\documentclass[a4paper,10pt,twoside]{cpc-hepnp}

\usepackage{multicol}
\usepackage{graphicx}
\usepackage{booktabs}
\usepackage{amssymb,bm,mathrsfs,bbm,amscd}
\usepackage[tbtags]{amsmath}
\usepackage{lastpage}

\begin{document}

\fancyhead[c]{\small Summit to Chinese Physics C } \fancyfoot[C]{\small 010201-\thepage}

\title{Simulation of the Capture and Acceleration Processes in HIAF-CRing}
\author{%
 SHANG Peng$^{1,2}$
\quad YIN Da-Yu$^{1}$\email{yindy@impcas.ac.cn}
\quad XIA Jia-Wen$^{1}$%
\quad YANG Jian-Cheng$^{1}$
\\ \quad QU Guo-Feng$^{1,2}$
\quad ZHENG Wen-Heng$^{1,2}$
\quad LI Zhong-Shan$^{1,2}$
\quad RUAN Shuang$^{1,2}$
}
\maketitle

\address{%
$^1$ Institute of Modern Physics, Chinese Academy of Sciences, Lanzhou 730000, China\\
$^2$ University of Chinese Academy of Sciences, Beijing 100049, China\\
}

\begin{abstract}
In order to investigate the longitudinal beam dynamics during the adiabatic capture and acceleration processes of the CRing in HIAF project, a simulation of both processes above is carried out with $U^{34+}$ ions. The ions will be captured into a bucket adiabatically and accelerated from 800 MeV/u to 1130 MeV/u . Simulation of these processes by tracking appropriate distributions with the longitudinal beam dynamics code ESME has been used to find optimum parameters such as RF phase, RF voltage and RF frequency etc.  An enhanced capture and acceleration efficiency can be gotten from the simulation results, with the optimized RF voltage and RF phase program.
\end{abstract}



\footnotetext[0]{\hspace*{-3mm}\raisebox{0.3ex}{$\scriptstyle\copyright$}2014
Chinese Physical Society and the Institute of High Energy Physics
of the Chinese Academy of Sciences and the Institute
of Modern Physics of the Chinese Academy of Sciences and IOP Publishing Ltd}%

\begin{multicols}{2}

\section{Introduction}

The HIAF (\textbf{H}igh \textbf{I}ntensity heavy ion \textbf{A}ccelerator \textbf{F}acility) project, which consists of high intensity ion source SERC, high intensity Heavy Ion Superconducting Linac (HISCL), a Booster Ring (BRing), a Compression Ring (CRing) and a multifunction storage ring system for heavy ion related researches, has been proposed by IMP (Institute of Modern Physics, Chinese Academy of Sciences) from 2009\cite{lab1}.
 As an important component of the HIAF complex, the synchrotron CRing has been designed to be a multi-function facility. The facility named after its core feature - beam compression - has a maximum magnetic rigidity of 43 Tm, and the basic
 parameters and its primary structure are shown in Table~\ref{ring_parameters} and Fig.~\ref{layout} respectively.

\begin{center}
\tabcaption{ \label{ring_parameters}  Machine parameters for CRing.}
\footnotesize
\begin{tabular*}{80mm}{l@{\extracolsep{\fill}}lr}
\toprule 
parameters               &signs\&units           & values   \\
\hline                                                      \\
Circumference            & $C_0$/m               & 804      \\
Average Radius           & $R$/m                 & 127.96    \\
Bending Radius           & $\rho$/m              & 19.11   \\
Max. Magnetic Induction  & $B_{max}$/T           & 2.25     \\
Max. Magnetic Rigidity   & $B\rho$/(Tm)          & 43       \\
Max. Ramping Rate        & $\dot{B}_{max}$/(T/s) & 1.125    \\
Initial Momentum Spread  & $\Delta{P}/P $        & $\pm5\times10^{-4}$         \\
Gamma Transition         & $\gamma_t$            &11.14          \\
\bottomrule
\end{tabular*}
\end{center}

The injected beam from BRing will firstly be accumulated into high current by a barrier bucket accumulation scheme, then the accumulated beam will be captured adiabatically into a stationary bucket to decrease the beam loss. Subsequently, CRing will accelerate the captured beam of heavy ions to a high energy, and compress the beam into a high intensity beam. The high intensity beam will be extracted for internal and external target experiments eventually.

\begin{center}
\includegraphics[width=7cm]{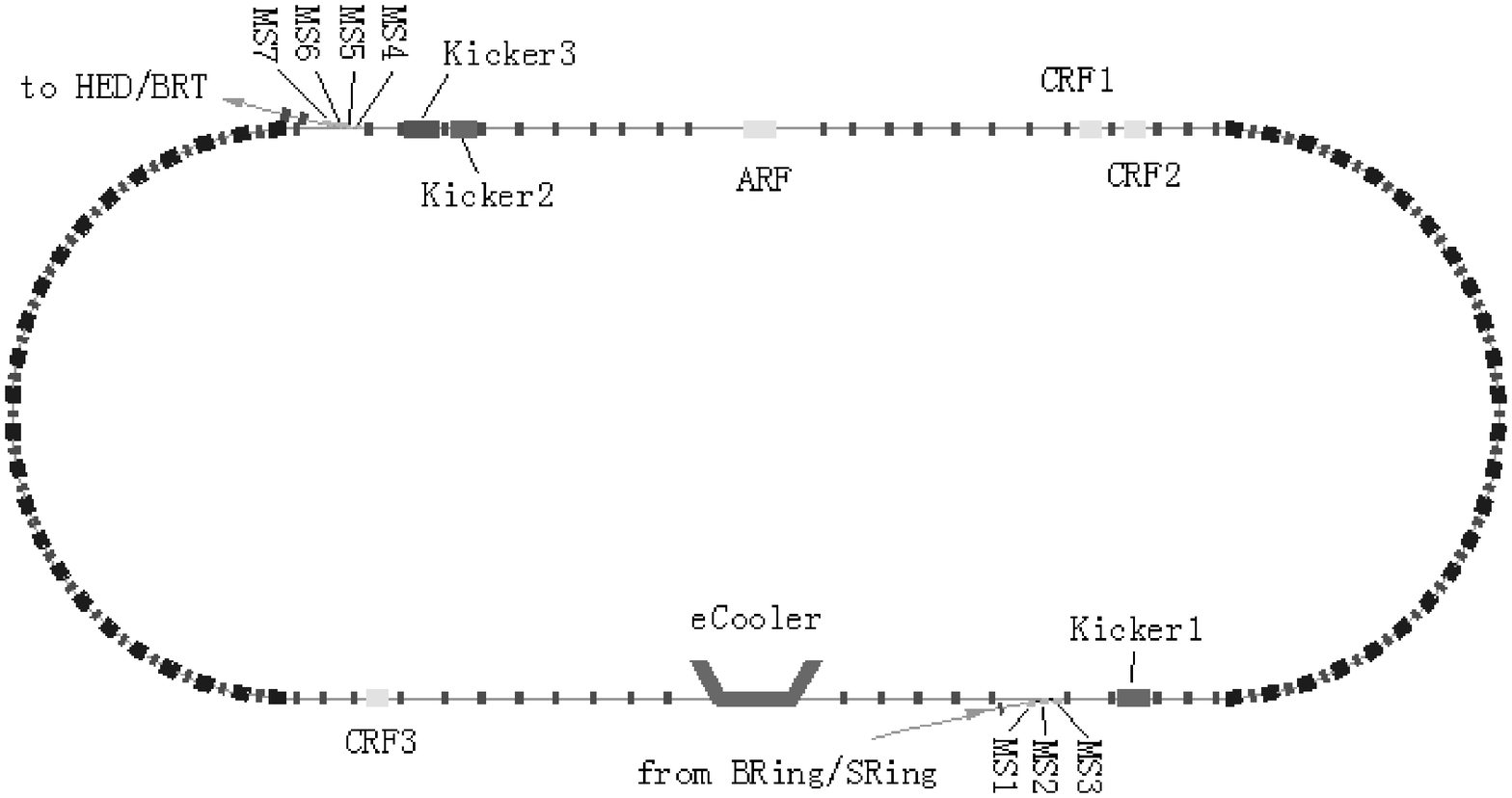}\\
\figcaption{\label{layout}   The layout of CRing. }
\end{center}

In order to get a high acceleration efficiency, the beam must be captured in a high efficiency first. In our simulations we generally considered a 800 MeV/u beam of $U^{34+}$ ions. Due to the limitation of magnetic rigidity, the typical beam can be accelerated to the energy of 1130 MeV/u. For convenience in describing we can divide our inquiry into two parts with details: An adiabatic capture process simulation and an acceleration process simulation.

The longitudinal beam dynamics program ESME, which has been developed to model those aspects of beam behavior in a synchrotron that are governed by the radio frequency systems, is used for computation. It follows the evolution of a macro-particle distribution in energy-phase coordinates by iterating a map corresponding to the single-particle equations of motion\cite{lab2}. 

The single particle equations for longitudinal motion in a synchrotron are naturally formulated as a pair of first order differential equations\cite{lab3}:

\begin{equation}
\label{differential}
  \left\{
   \begin{aligned}
   \frac{\mathrm{d} }{\mathrm{d} t}\left ( \frac{\Delta E}{\omega_s} \right ) &=
   \frac{q}{2\mathrm{\pi}}\left [ V(\varphi)- V(\varphi_s) \right ]  \\
   \frac{\mathrm{d} }{\mathrm{d} t}\Delta \varphi &=
   -\frac{h \omega_s \eta}{\beta^{2} E}\Delta E \\
   \end{aligned}
   \right.
   .
  \end{equation}

Eqs.~(\ref{differential}) are used as a basic theory in computation of the simulation. Here $\Delta E = E - E_s$, $\Delta \varphi = \varphi - \varphi_s$, $\eta = \frac{1}{\gamma^{2}}-\frac{1}{\gamma_t^{2}}$ with $\gamma_t$ the Lorentz factor at transition energy. $E$ and $\varphi$ refer the total energy and the phase of the particle. $V(\varphi) = V_{rf} \cdot \sin(\varphi)$, where $V_{rf}$ is the RF voltage. $\omega$ is the frequency and $h$ is the harmonic number, in this paper $h=1$. The suffix "s" refers to the synchronous particle.

\section{Adiabatic capture}
The coasting beam accumulated in CRing is captured using the adiabatic program in this simulation. The initial distribution of coasting beam is random uniform in phase $\varphi$, Gaussian in energy $E$ in the phase space coordinate, see in Fig.~\ref{initial_dst}. The process is called adiabatic when the relative change in bucket height or bucket area is much slower than the synchrotron frequency\cite{lab4}.

\begin{center}
\includegraphics[width=7cm]{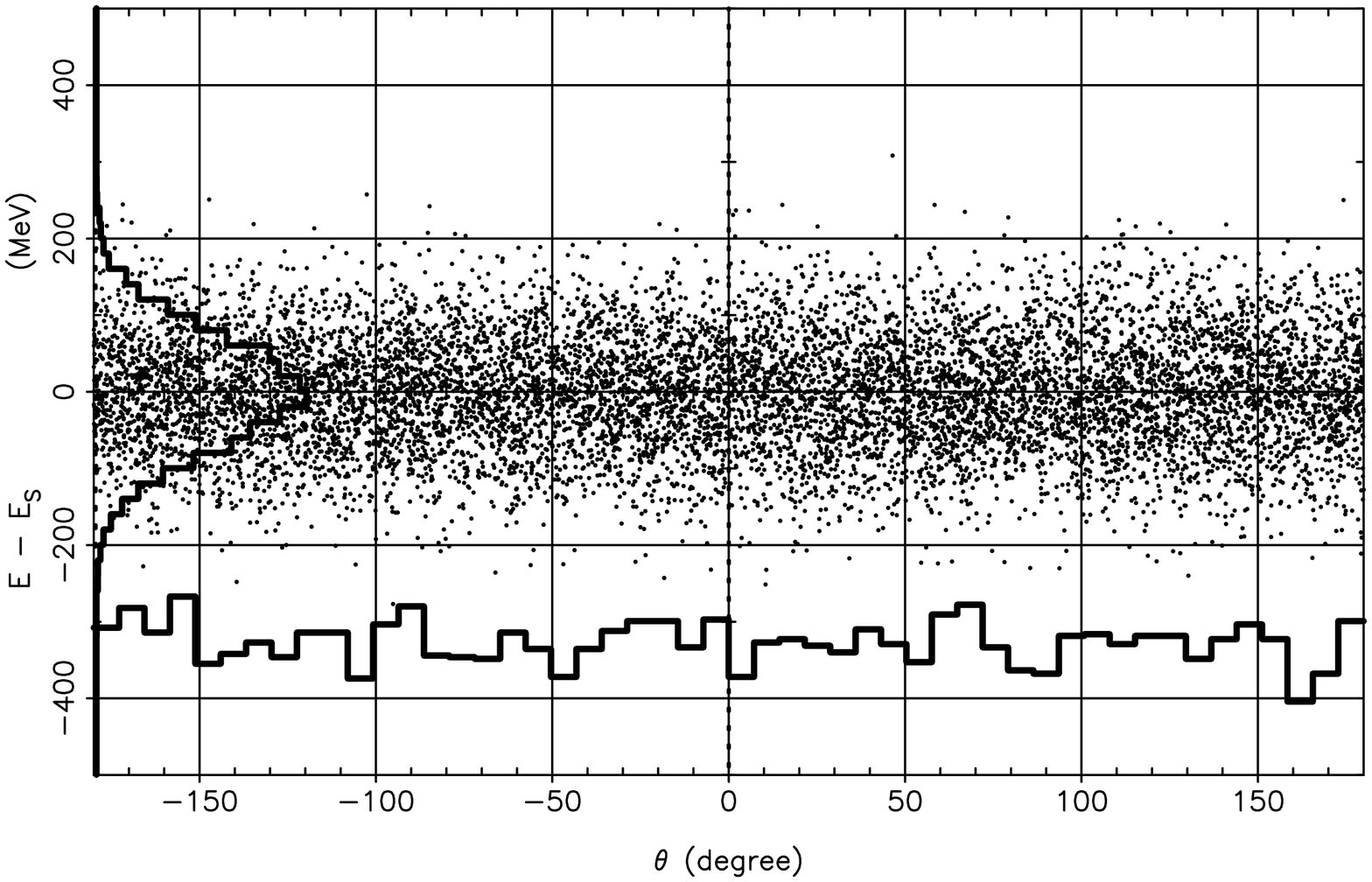}
\figcaption{Initial distribution in phase-energy space.}
\label{initial_dst}
\end{center}

The parameters that define the adiabatic process are chosen in order to satisfy three criteria. The capture efficiency ¨C the ratio between injected and captured particles - has to be a certain value, the dilution ¨C the longitudinal emittance increase ¨C has to be minimum, the bunching factor ¨C the ratio between the phase length of the bunch and the RF wavelength ¨C has to be maximum to reduce space charge forces\cite{lab5}.

To satisfy these criteria, the adiabatic factor $n_{ad}$\cite{lab6}, the bucket-to-bunch area ratio $A_B/A_b$ ,and the ratio between the final and initial cavity voltage $V_f/V_i$  have been fixed by following equations:

\begin{equation}
\label{eq2}
n_{ad} = \frac{\omega_{si} t}{1- \sqrt{\frac{V_i}{V(t)}}}  .
\end{equation}

\begin{equation}
\label{eq3}
     A_B/A_b=\frac{3}{2}     .
\end{equation}

\begin{equation}
\label{eq3-1}
     V_i/V_f=\frac{1}{30}       .
\end{equation}

Where $\omega_{si}$ is the initial angular frequency of synchronous particle. $V_i$ and $V_f$ are the initial and final cavity voltage of the capture process respectively, $V(t)$ is the function of the RF voltage increasing with time, $A_B$ is the bucket area, and $A_b$ is the bunch area.

$V_f$ and $A_b$ can be derived according to the following formulas:
\begin{equation}
\label{A_b}
     A_b = 2\mathrm{\pi}\times2\frac{\Delta E}{\omega_{si}}     .
\end{equation}

\begin{equation}
\label{V_cap}
      V_f = \left(\frac{A_B}{16}\right)^2\frac{2\mathrm{\pi}h|\eta|\omega_s^2}{q\beta^2 E_s} .
\end{equation}

These choices completely fix the capture process. We calculated a time-varying adiabatic program of voltage applied to capture process according to relationships mentioned above,see in Fig.~\ref{Vcap}. In this process, other time-varying programs such as magnetic field, RF frequency 
keep constant in consequence the beam energy doesn't change. The RF phase stays in zero all the time, providing a perfect matching of the stationary bucket for the accumulated beam. These two features above are fundamental differences between capture process and acceleration process.

\begin{center}
\includegraphics[width=7cm]{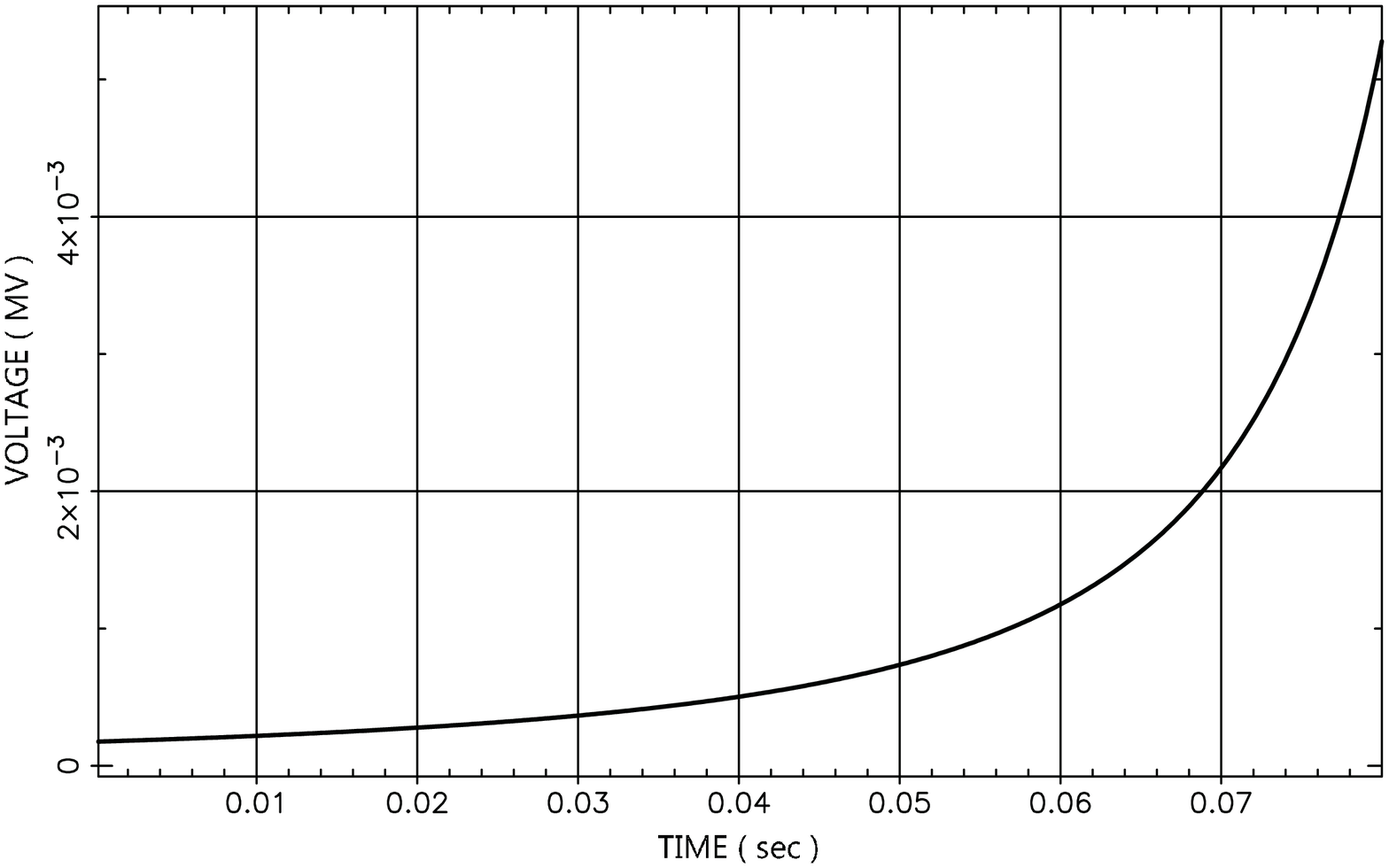}
\figcaption{The adiabatic capture program of voltage.}
\label{Vcap}
\end{center}

Table~\ref{tab2} presents the relevant parameters using in capture process. The capture time seems to be very long because of the very large synchrotron period , which is
\begin{equation}
\label{sychrotron_period}
      T_s = \tau_0\sqrt{\frac{2\mathrm{\pi}\beta^2E}{eV_0~h\eta}} .
\end{equation}
at injection energy and 5.284 kV capture voltage, here $\tau_0$ is the circulating period about several microseconds in general.

\begin{center}
\tabcaption{ \label{tab2}  Adiabatic capture, relevant parameters.}
\footnotesize
\begin{tabular*}{85mm}{l@{\extracolsep{\fill}}lr}
\toprule 
paremeters                 &signs\&units           & values    \\
\hline                                                         \\
Num. of particles          & $N$                   & 10000     \\
Capturing Energy           & $E_i$/(MeV/u)         & 800       \\
Bunch Area                 & $A_b$/($eVs$)         & 931.74    \\
Bucket Area                & $A_B$/($eVs$)         & 1397.61   \\
Initial Capturing Voltage     & $V_i$/kV              & 0.176     \\
Final Capturing Voltage       & $V_f$/kV              & 5.284        \\
Capture Time               & $t_{cap}$/(ms)        & 80    \\
Adiabatic Parameter        & $n_{ad} $/m           & 2.0$\times10^5$    \\
Synchrotron Period\cite{lab7}         & $T_s$/($\mu s$)       & 19.20     \\
Filling Factor             & $A_b/A_B$             & 0.666   \\
Capture Field              & $B_{cap}$/T            & 1.7835         \\
\bottomrule
\end{tabular*}
\end{center}

The RF voltage program for adiabatic capture is given\cite{lab4} in equation (\ref{V_adiabatic_capture}).

\begin{equation}
\label{V_adiabatic_capture}
V(t)=\frac{V_i}{\left[ 1- \left( 1- \sqrt \frac{V_i}{V_f} \right)\frac{t}{t_{cap}} \right]^2}   .
\end{equation}

The simulation gives a theoretical 100\% efficiency as shown in Fig.~\ref{cap_process}-(f), which means all 10000 particles exist in the whole capture process, and are bounded in separatrix formed by RF voltage and RF phase at the end of this process.

\begin{center}
\begin{multicols}{2}
\centering
\includegraphics[width=4cm]{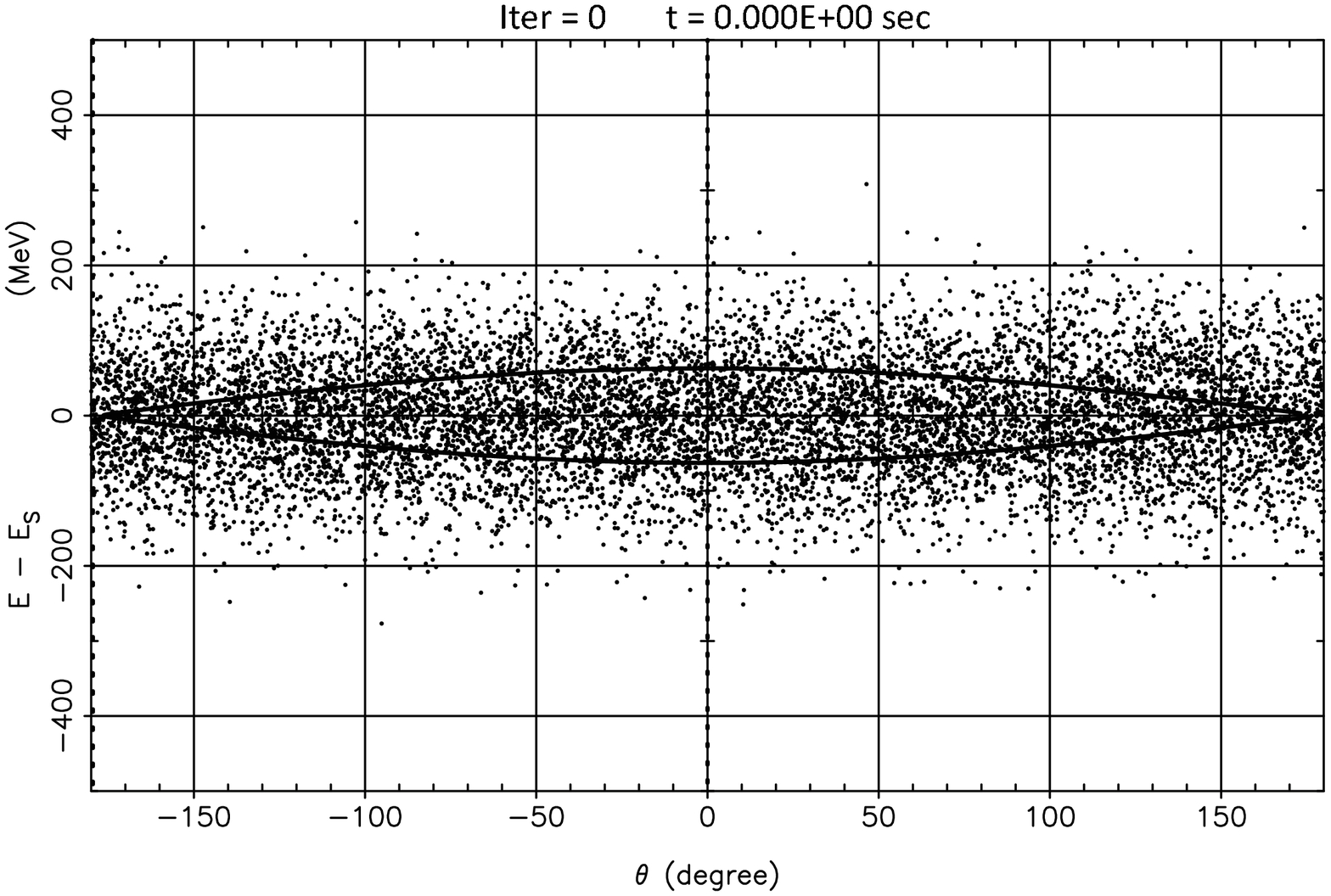}
\\ \centering \footnotesize (a)  \\
\includegraphics[width=4cm]{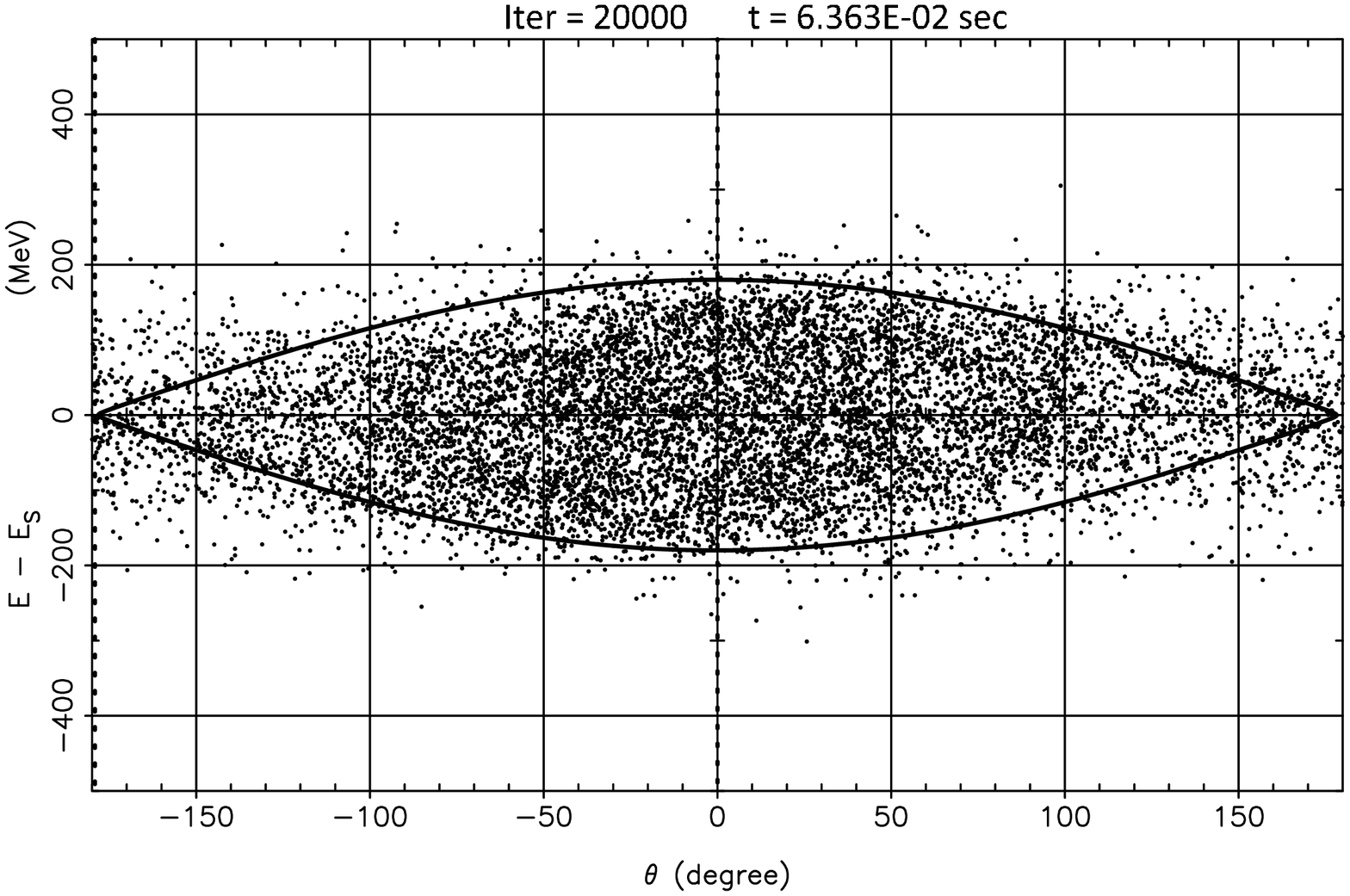}
\\ \centering \footnotesize (b)  \\
\includegraphics[width=4cm]{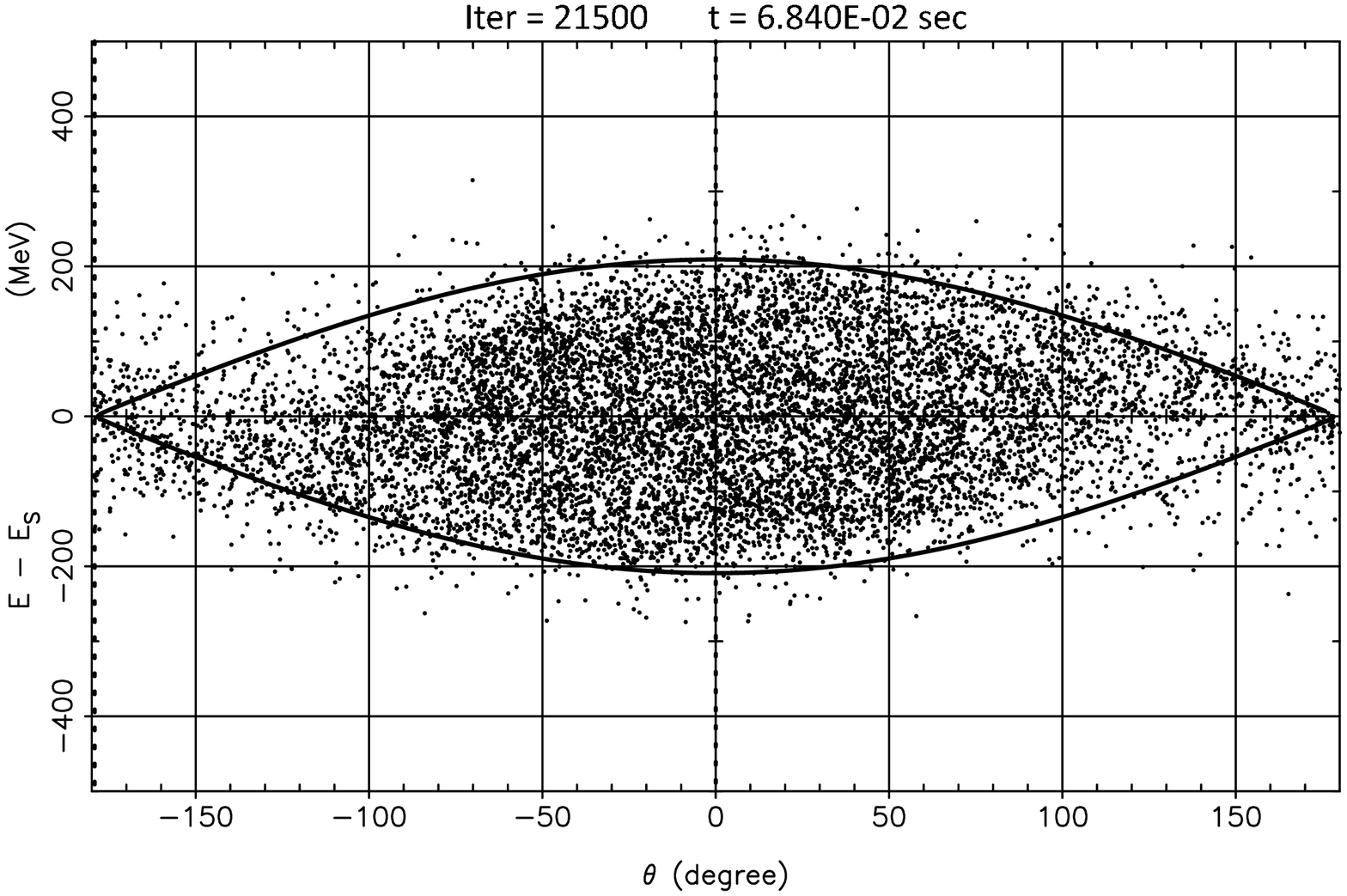}
\\ \centering \footnotesize (c)   \\
\includegraphics[width=4cm]{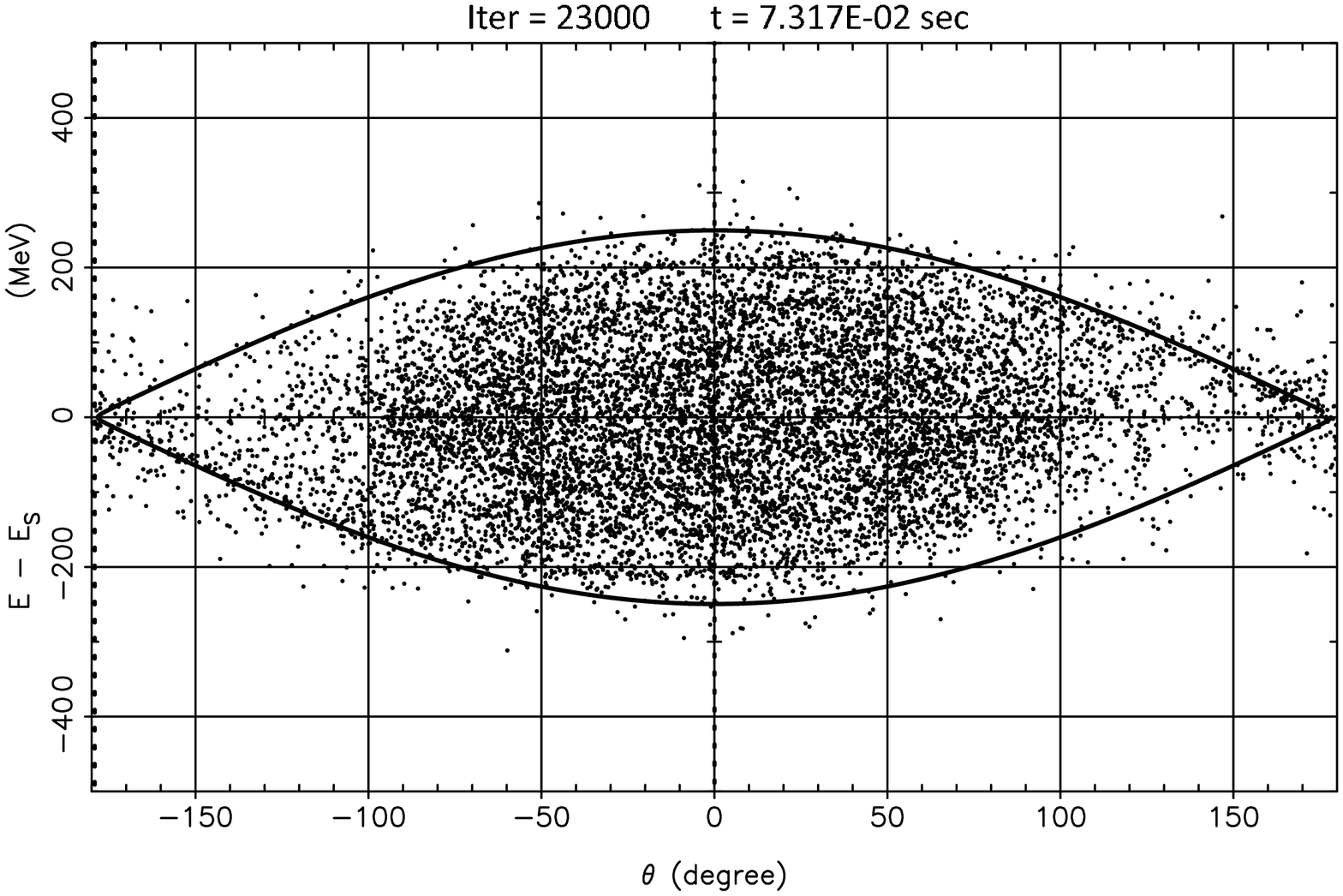}
\\ \centering \footnotesize (d)   \\
\includegraphics[width=4cm]{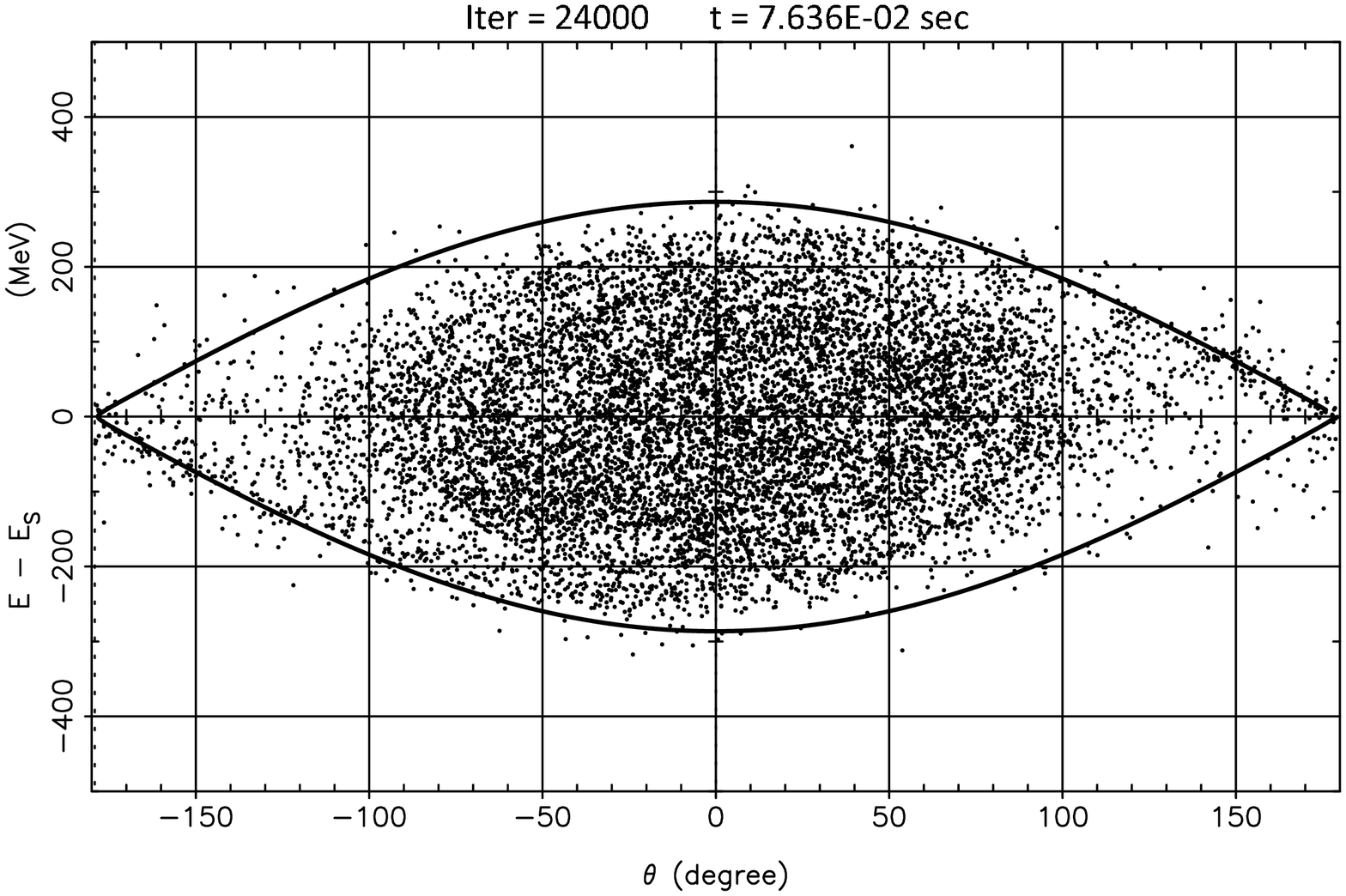}
\\ \centering \footnotesize (e)   \\
\includegraphics[width=4cm]{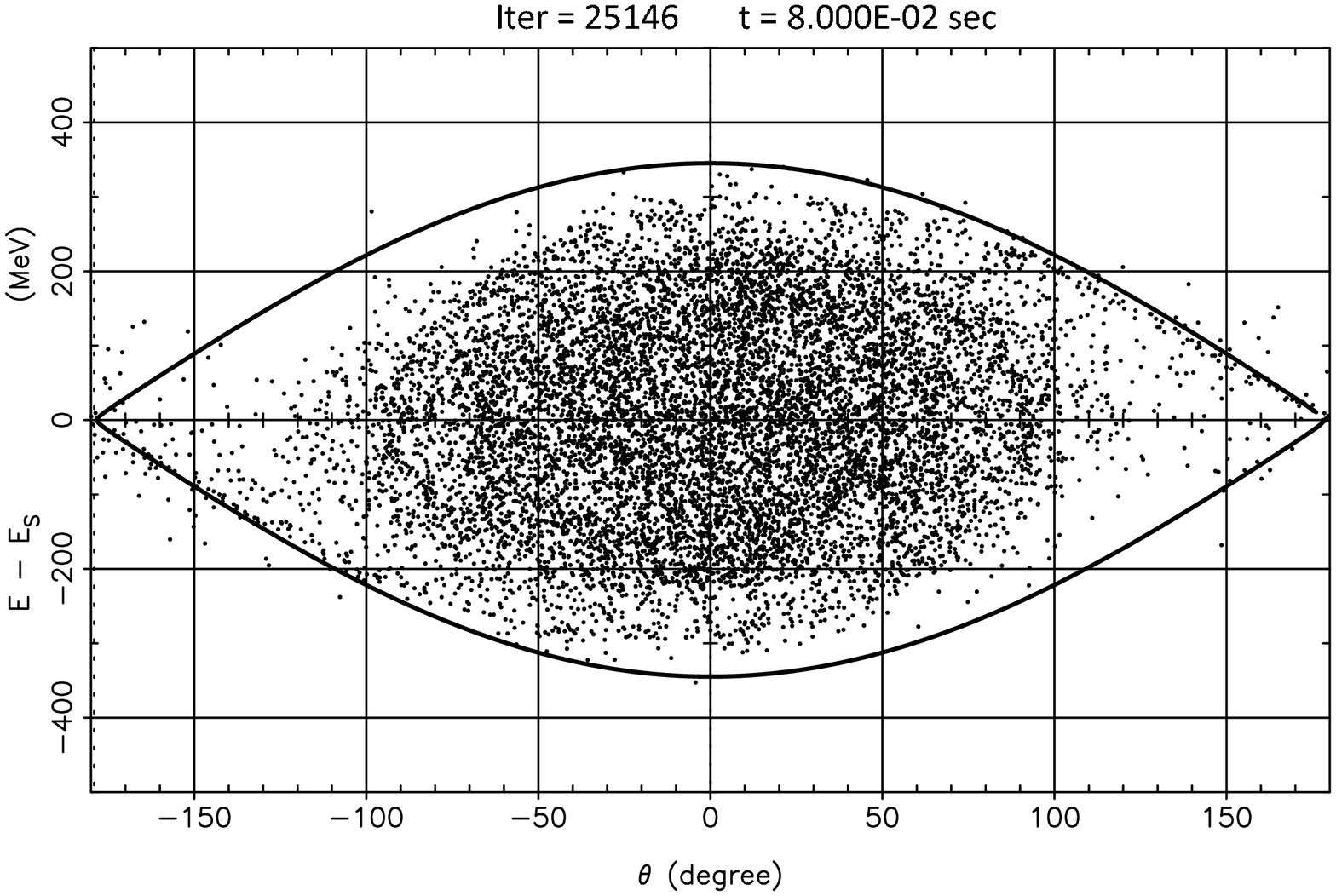}
\\ \centering \footnotesize (f)   \\

\end{multicols}
\figcaption{ Particle distribution of adiabatic capture using ESME program at the beginning i.e. 0th turn, 20000th turn, 21500th turn, 23000th turn, 24000th turn and 25146th turn which is the end of the capture process.}
\label{cap_process}
\end{center}

\section{Acceleration process}

After adiabatic capture process, we got a Gaussian like distribution in both $E$ and $\varphi$ directions, see in Fig.~\ref{begin_acc}. The coasting beam have been bunched into a bucket. We keep the harmonic number $h=1$ at the beginning of acceleration.

This section describes the basic acceleration process and simulation results of acceleration. The energy of synchronous particle will be accelerated from 800 MeV/u to 1130 MeV/u.

\begin{center}
\includegraphics[width=7cm]{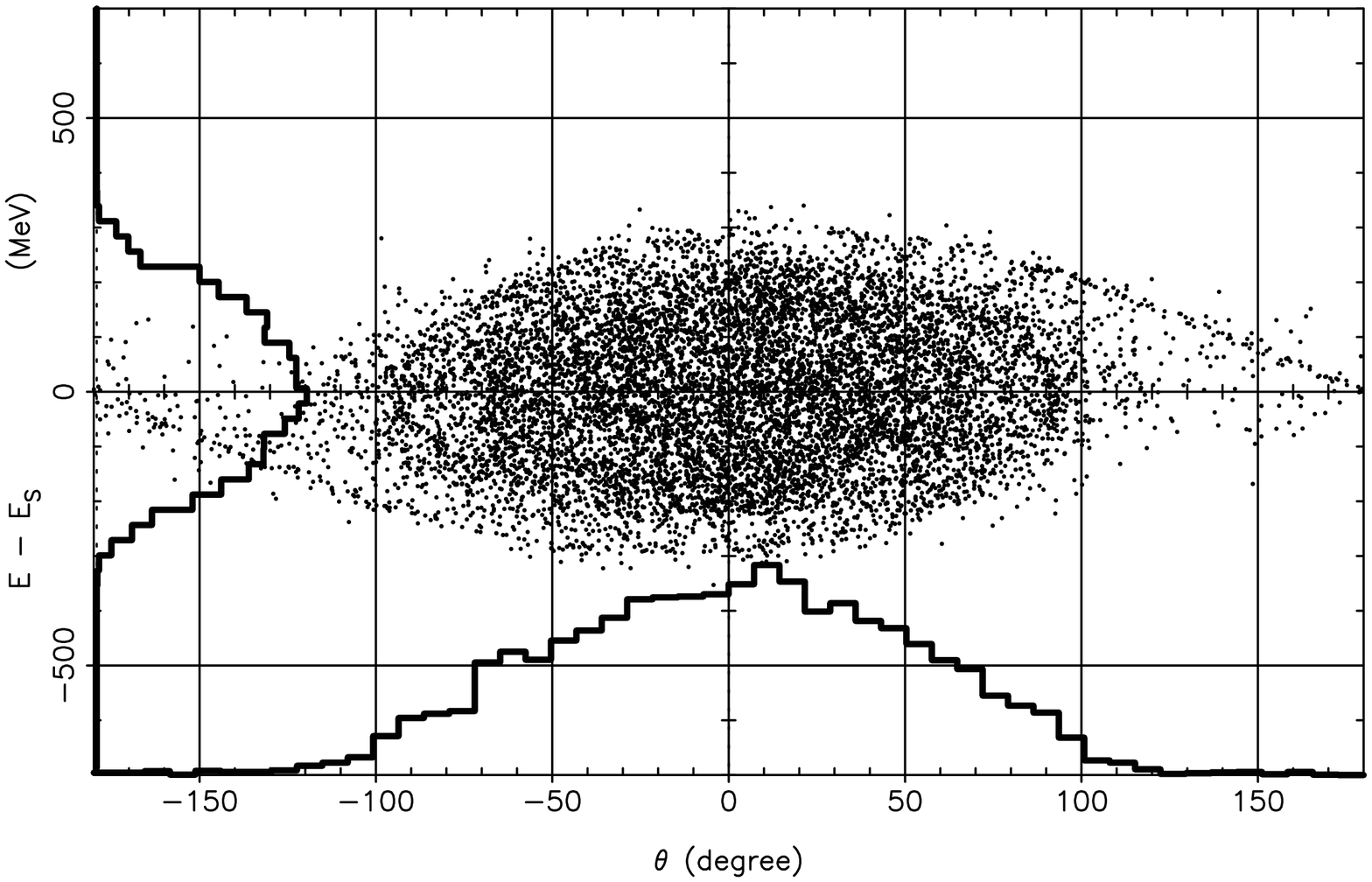}
\figcaption{Particle distribution at the beginning of acceleration.}
\label{begin_acc}
\end{center}

\subsection{Acceleration process description}
Ingeneral, particles gain energy from the electric field in the longitudinal direction.
The particle energy must synchronize with the magnetic field generated by dipole strictly to keep the synchronous particle moving on a closed orbit.

We have mentioned the longitudinal motion equations before, in which we assumed that the RF electric field is distributed along CRing uniformly. While in fact the RF cavity is located in a straight section of the ring. So we use the difference equations (\ref{difference})\cite{lab8} to supersede the Eqs.~(\ref{differential}), which means we calculate the energy and phase of particles turn by turn. One equation gives the phase slip between a particle and a synchronous particle during the passage between RF gaps, and one gives the energy change at the gap:

\begin{equation}
\label{difference}
  \left\{
   \begin{aligned}
   \delta_{n+1} &=\delta_{n}+\frac{q}{\beta^2 E_s}\left [ V(\varphi)- V(\varphi_s) \right ]
        \\
   \varphi_{n+1} &= \varphi_n + 2 \mathrm{\pi}h\eta \delta_{n+1}\\
   \end{aligned}
   \right.
   .
  \end{equation}

   The RF voltage was determined by two criteria:
  \begin{enumerate}
    \item 	The energy gain of a particle must match the increase of bending magnetic field, in other words, the RF voltage $V_{rf}$ have to satisfy the relationship below\cite{lab9}:
       \begin{equation}
       \label{sychronous_condiction_1}
        V_{rf}\sin(\varphi_s) = 2\mathrm{\pi}\rho R \dot B    .
       \end{equation}
    $\dot B$ is the ramping rate, i.e. bending magnetic field changing with time.

    \item 	The RF bucket area $A_B$ has to be greater than or equal to the longitudinal beam emittance $A_b$. In this paper, we set the relationship that
       \begin{equation}
       \label{filling_factor}
        A_B/A_b =\frac{3}{2}    .
       \end{equation}
    as same as the capture process.
    This is not the only way to provide an acceleration bucket which is large enough to bound most particles. Other schemes can be effective in acceleration too.(see reference \cite{lab5})
  \end{enumerate}

  Let us assume that $\dot R = 0$. According to Eq.~(\ref{sychronous_condiction_1}), we see that $V_{rf} \sin(\varphi_s )$ scales with $\dot B$ only.

We can easily get the RF bucket area $A_B$ if we know the longitudinal beam emittance $A_b$ which we can calculate with a uniform distribution at injection, see in Eq.~(\ref{A_b}).

We can achieve another relationship between $V_{rf}$ and $\varphi_s$ in $(\varphi,\Delta E/ \omega_s)$ phase space using the formula~(\ref{eq8}).
 \begin{equation}
        A_B = 16\sqrt{\frac{qV_{rf}\beta^2E_s}{2 \mathrm{\pi}h|\eta|\omega_s^2}}\alpha_b(\varphi_s) \label{eq8}   .
 \end{equation}

 Here $\alpha_b(\varphi_s)$ is the ratio of the bucket area between a running bucket $(\varphi_s \neq 0)$ and a stationary bucket $(\varphi_s = 0)$. Note that $\alpha_b(\varphi_s)$ can be approximated by a simple function:

 \begin{equation}
 \alpha_b(\varphi_s)\approx \frac{1-\sin{\varphi_s}}{1+\sin{\varphi_s}} \label{eq9}   .
\end{equation}

Then we have the equation:
\begin{equation}
 A_B = 16\sqrt{\frac{qV_{rf}\beta^2E_s}{2 \mathrm{\pi}h|\eta|\omega_s^2}}
 \left[\frac{1-\sin{\varphi_s}}{1+\sin{\varphi_s}}\right] \label{sychronous_condiction_2}    .
\end{equation}

Every moment we know $\dot B$ and $A_B$ during acceleration, we can calculate the parameters of the synchronous particle such as $E_s, \omega_s, \beta$ and $\eta$. Then we can derive a simple cubic equation in either $V_{rf}$ or $\varphi_s$ by solving Eqs.~(\ref{sychronous_condiction_1}) and~(\ref{sychronous_condiction_2})\cite{lab7}. A small program was built in Fortran to get the $V_{rf}$or $\varphi_s$ by giving $B(t)$and $A_B$ as a matter of convenience during simulation.

\subsection{Simulation process}

The simulation requires magnetic field program $B(t)$, RF voltage program $V_{rf}(t)$ and Phase program $\varphi_s(t)$ for the acceleration process in ESME. We shall assume that the magnetic field changes with time as follows:

\begin{equation}
\label{magnetic_function}
\left\{
    \begin{aligned}
   B(t) &= B_{cap}+a(t-t_{cap})^3          \\
   \dot B(t) &= 3 a (t-t_{cap})^2          \\
   \end{aligned}
\right.
.
\end{equation}

The parameter $a$ is chosen when $\dot B$ reaches the maximum value $\dot B^\ast$ at time $t^\ast$. Thus
\begin{equation}
\label{eq12}
    a =\frac{\dot B^\ast}{3(t^\ast-t_{cap})^2}    .
\end{equation}

Table~\ref{tab3} shows the relevant parameters in acceleration process. From the table we notice that the RF voltage at the start of acceleration is the same value as the final capturing voltage. By this arrangement, we can ensure that the RF voltage changes smoothly without breaking in the actual implementation.

\begin{center}
\tabcaption{ \label{tab3}  Acceleration, relevant parameters.}
\footnotesize
\begin{tabular*}{85mm}{l@{\extracolsep{\fill}}lr}
\toprule 
paremeters                 &signs\&units           & values   \\
\hline                                                      \\
Num. of particles          & $N$                   & 10000   \\
Acceleration Duration      & $t_{acc}$/(ms)         &446           \\
Max. Value of $\dot B$     & $\dot B^\ast$/(T/s)    &1.125            \\
Corresponding time point   & $t^\ast$/(ms)          &50            \\
Corresponding RF Voltage   & $V^\ast$/(kV)          &37.7            \\
Corresponding Phase        & $\varphi_s^\ast$/(deg) & 27.3           \\
Voltage at the start of this process & $V_{start}$/(kV)  & 5.284            \\
Voltage at the end of this process & $V_{end}$/(kV)  & 32.7            \\
Coefficient $a$            &  $T/(s^3)$             &150 \\
\bottomrule
\end{tabular*}
\end{center}

Fig.~\ref{acc_curves} presents the variation tendency of magnetic field $B(t)$, as well as RF voltage $V_{rf}(t)$, synchronous phase $\varphi_s(t)$ and RF frequency $f_{rf}$. We see that the RF voltage got a maximum value when $B(t)$ reaches the maximum value $\dot B^\ast$.

\begin{center}
\begin{multicols}{2}
\centering
\includegraphics[width=4cm]{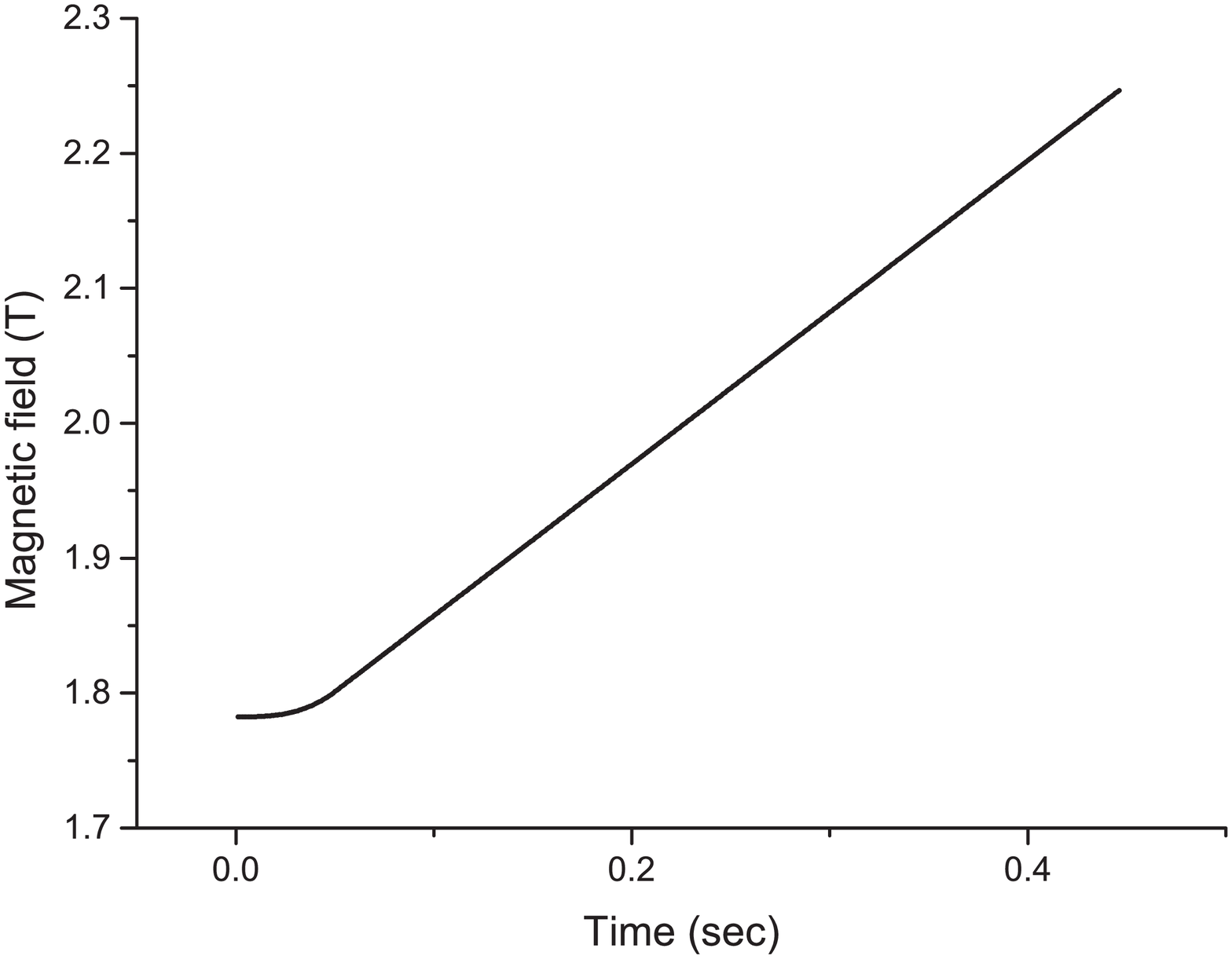}
\\ \centering \footnotesize (a)   \\
\includegraphics[width=4cm]{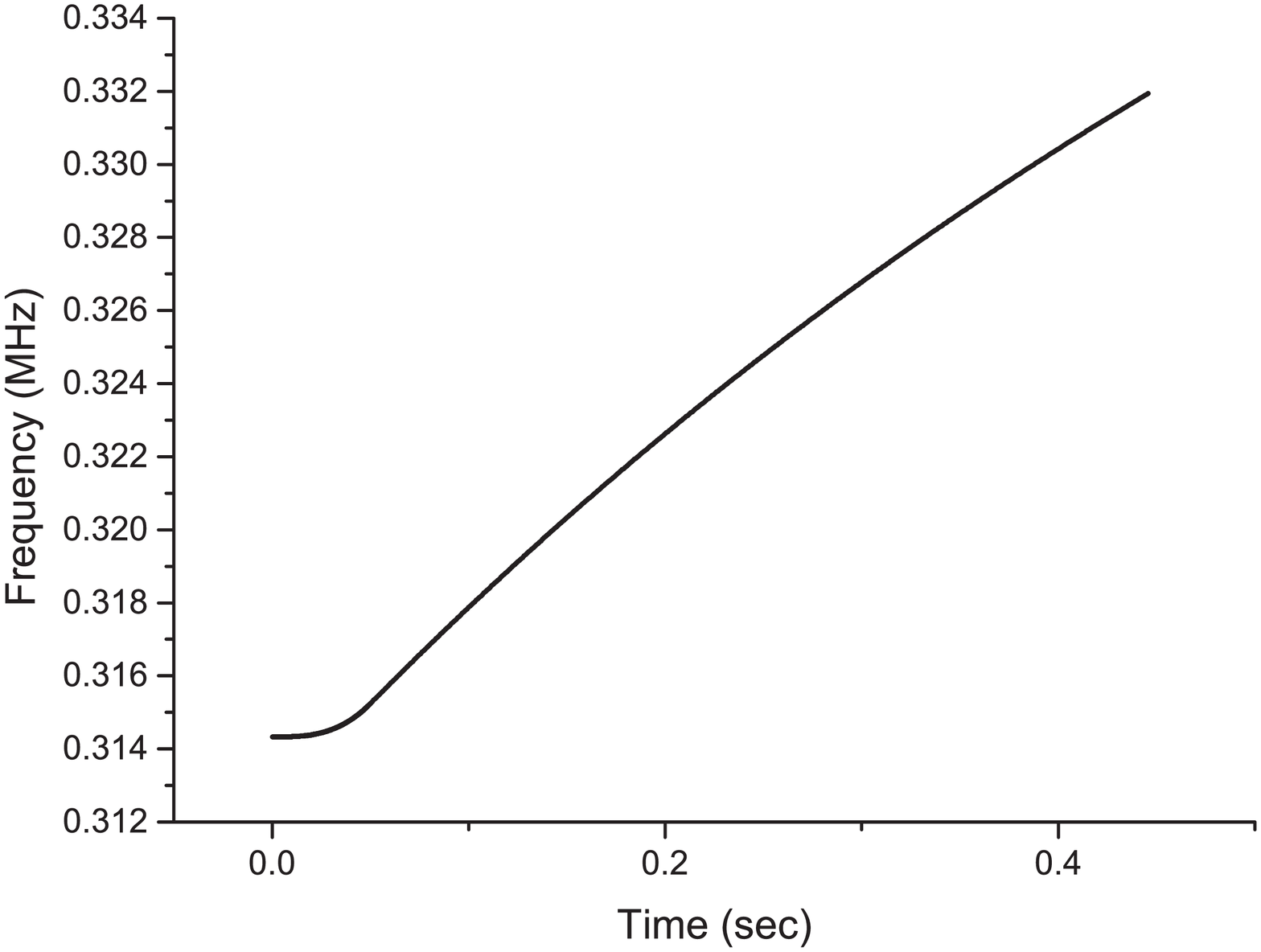}
\\ \centering \footnotesize (c)   \\
\includegraphics[width=4cm]{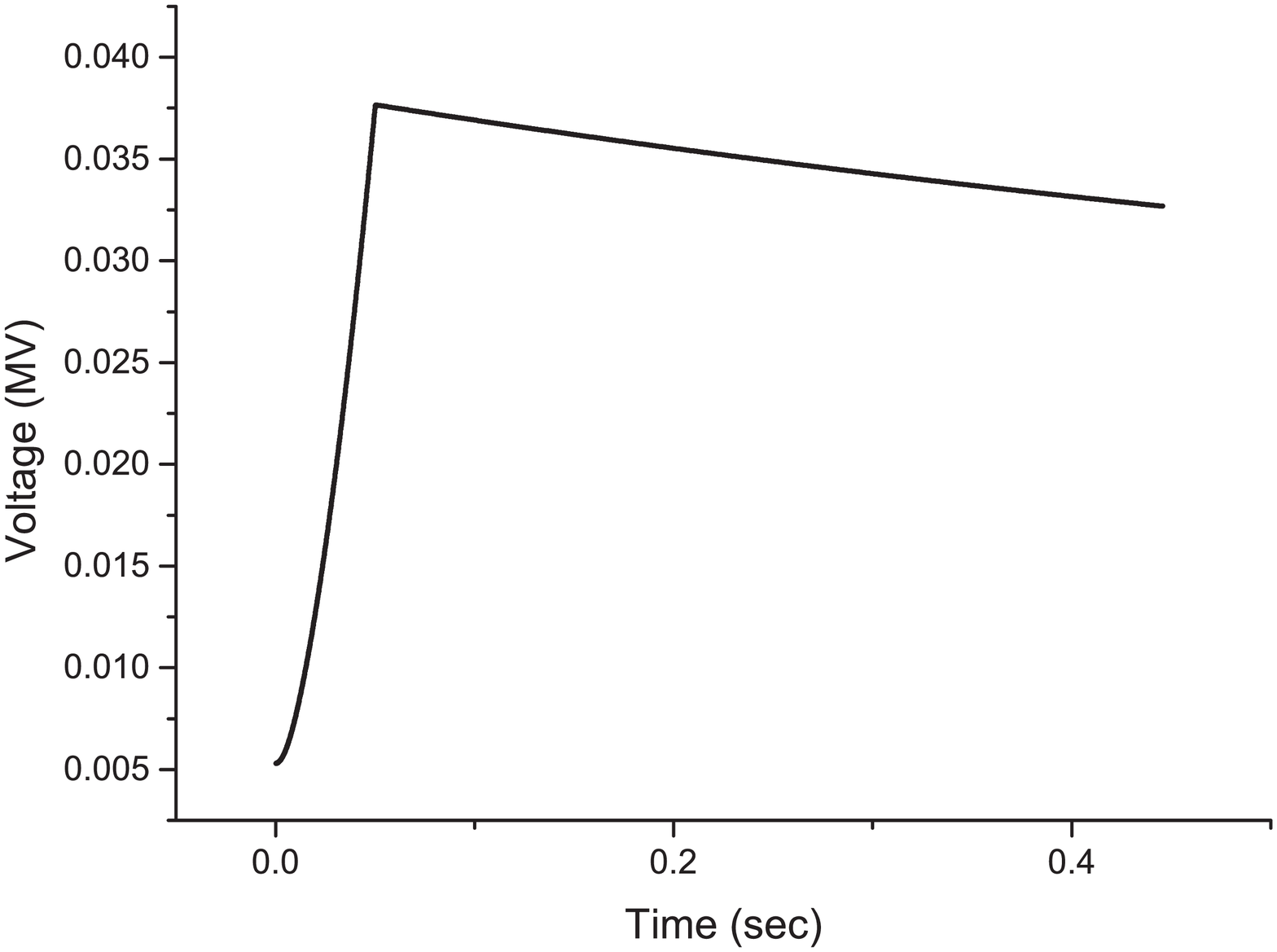}
\\ \centering \footnotesize (b)   \\
\includegraphics[width=4cm]{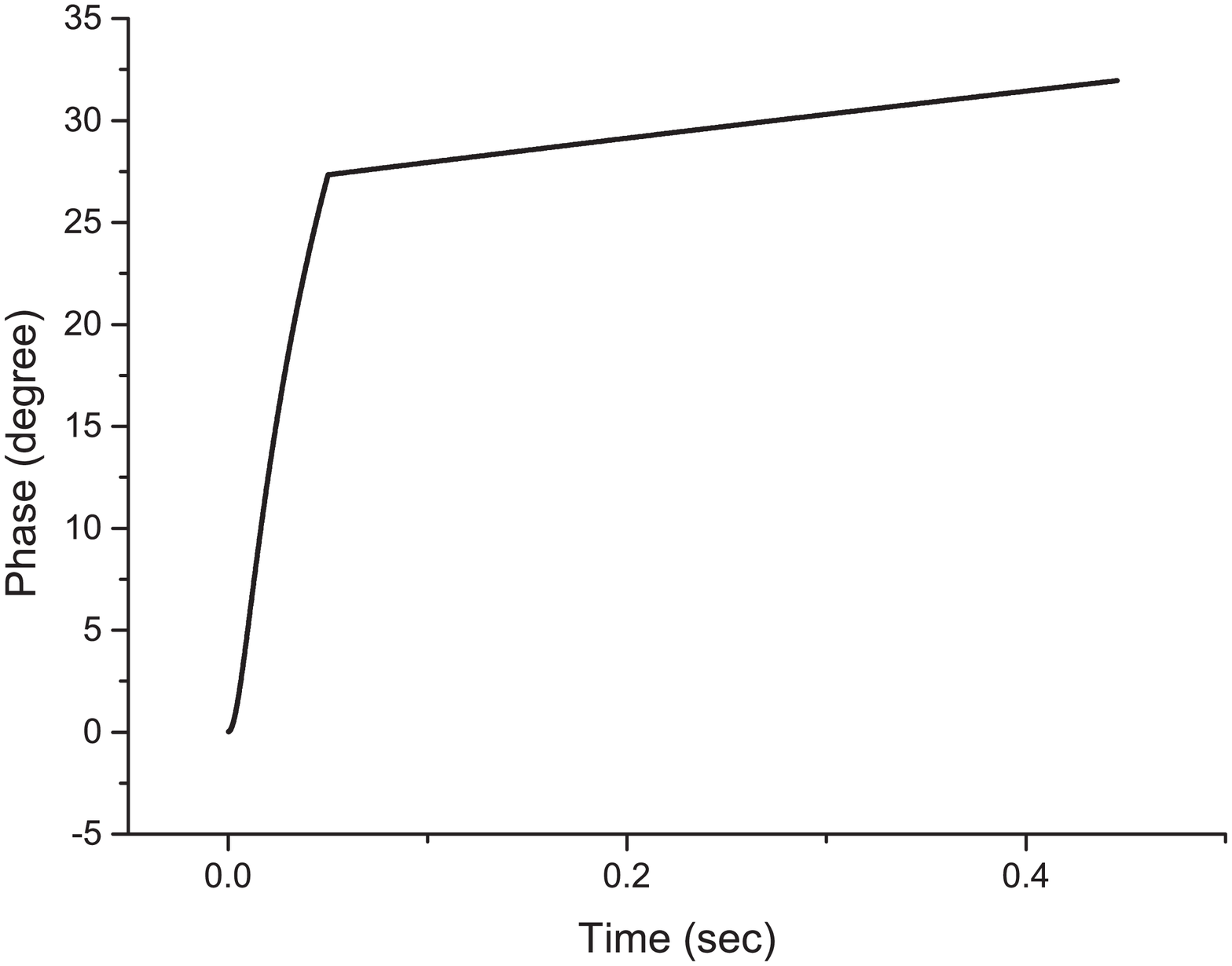}
\\ \centering \footnotesize (d)   \\
\end{multicols}
\figcaption{Parameter programs used in acceleration process.}
\label{acc_curves}
\end{center}

The simulation results of this process are shown in Fig.~\ref{acc_process}  utilizing the programs mentioned above. Compare Fig.~\ref{acc_process}-[(a)-(c)] to Fig.~\ref{acc_process}-[(d)-(f)], we notice that the bucket changes faster before $t^\ast$ when the ramping rate reaches its maximum value $\dot B^\ast$. This is concordant with Fig.~\ref{acc_curves}. After about 446 ms acceleration, the beam was accelerated up to 1130 MeV/u successfully. There are 9925 particles among the 10000 particles injected survived at the end of acceleration according to the simulation result. So we know the acceleration efficiency is about 99.3\%, which is pretty high compared with other acceleration methods.

\begin{center}
\begin{multicols}{2}
\centering
\includegraphics[width=4cm]{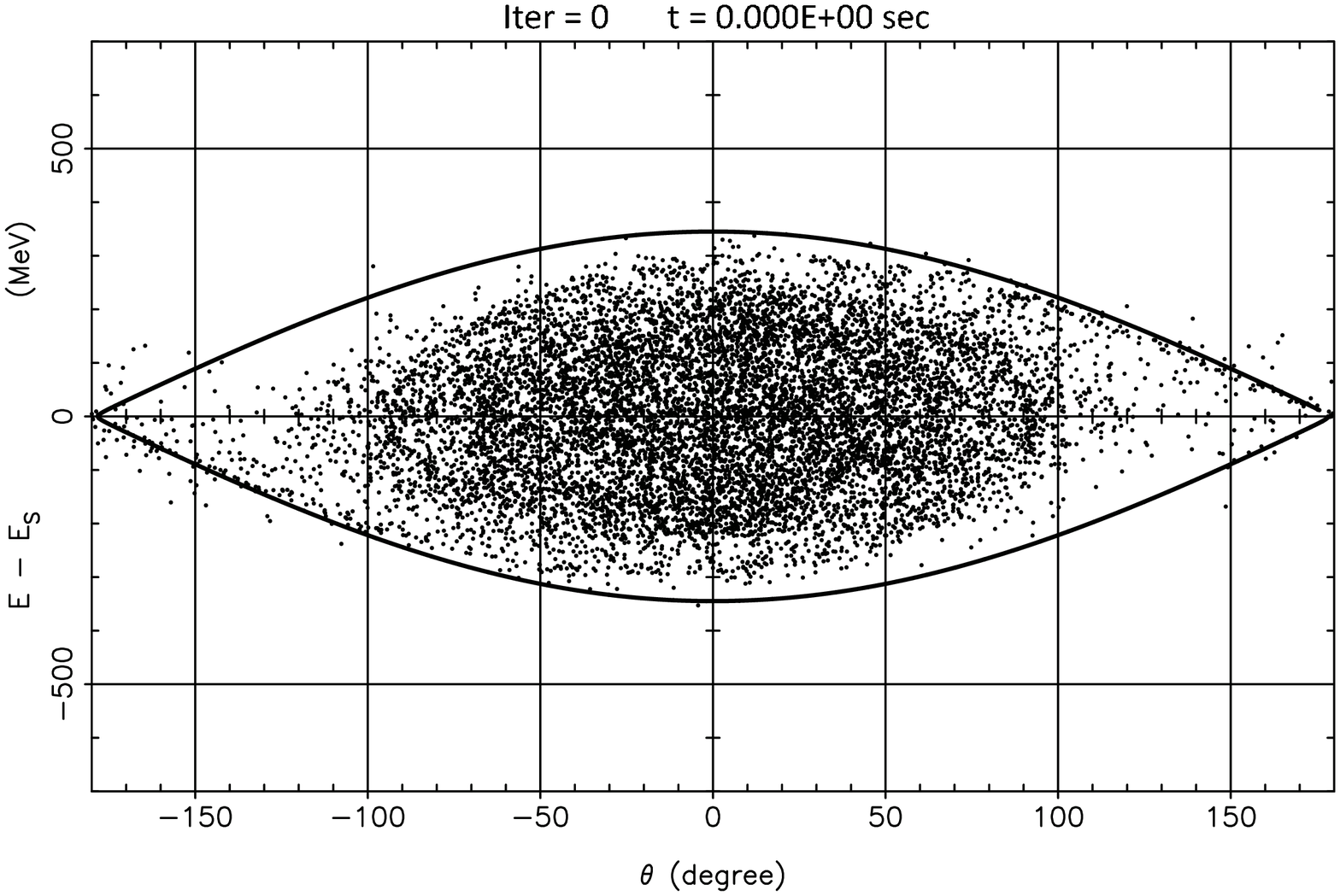}
\\ \centering \footnotesize (a)   \\
\includegraphics[width=4cm]{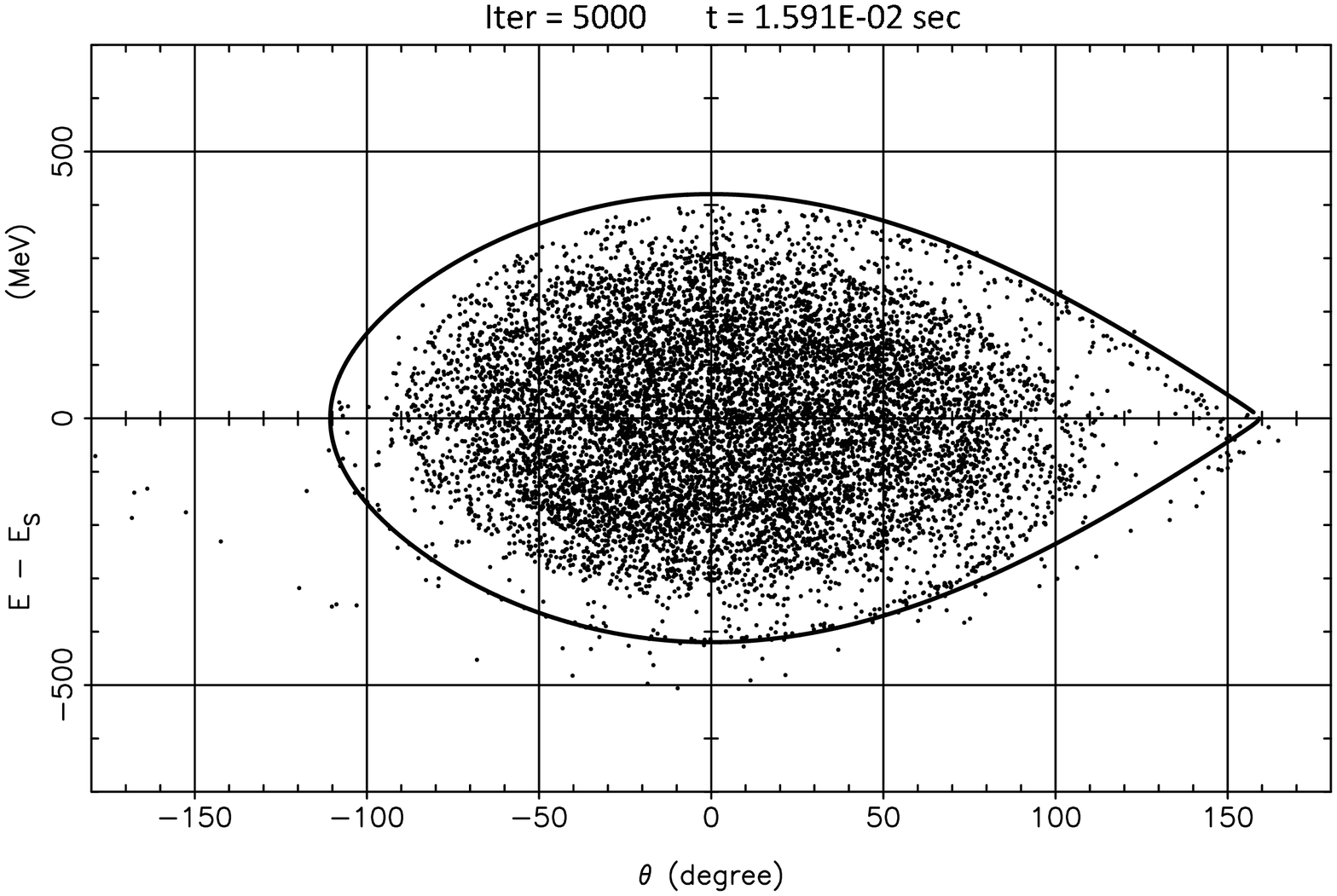}
\\ \centering \footnotesize (b)  \\
\includegraphics[width=4cm]{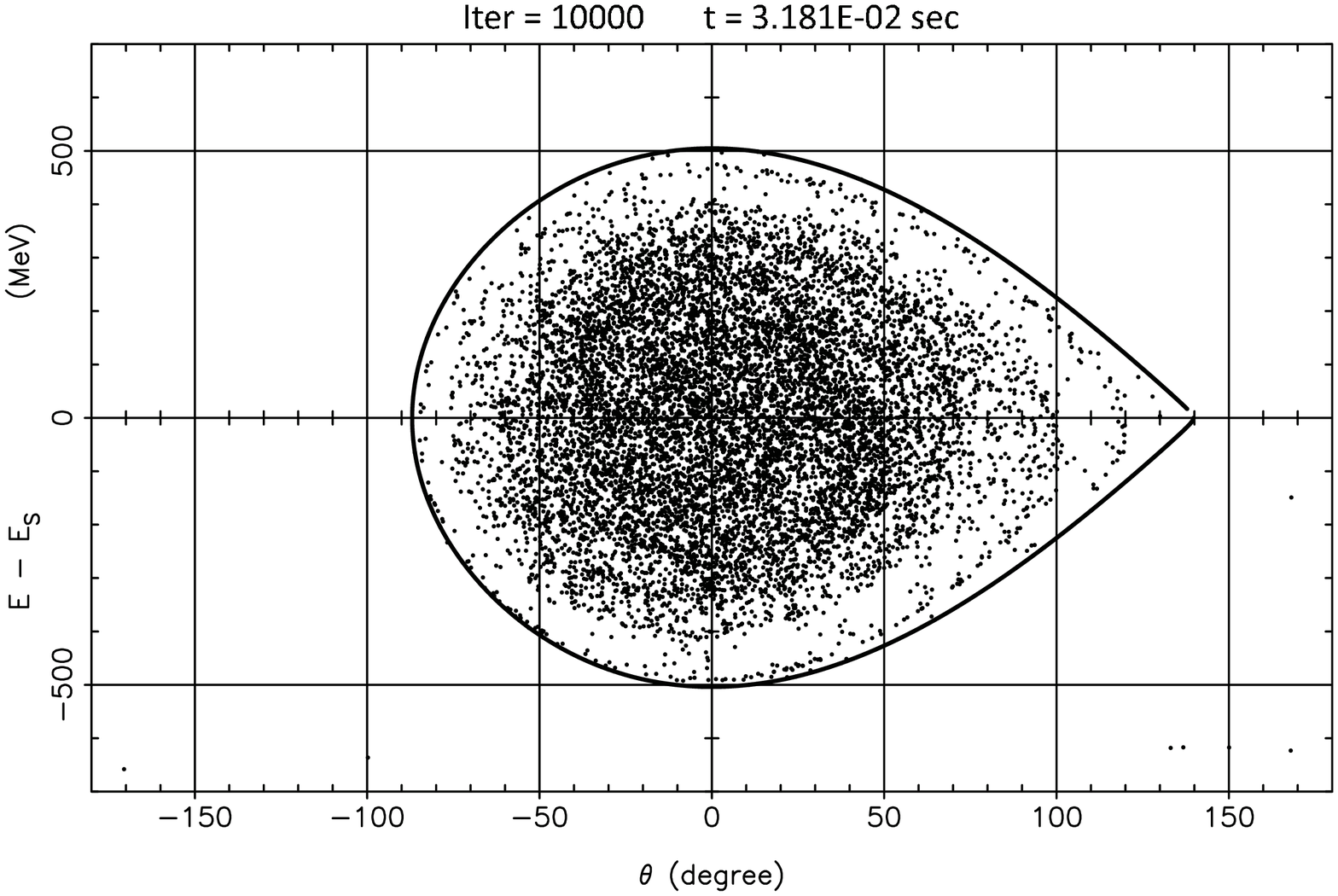}
\\ \centering \footnotesize (c)   \\
\includegraphics[width=4cm]{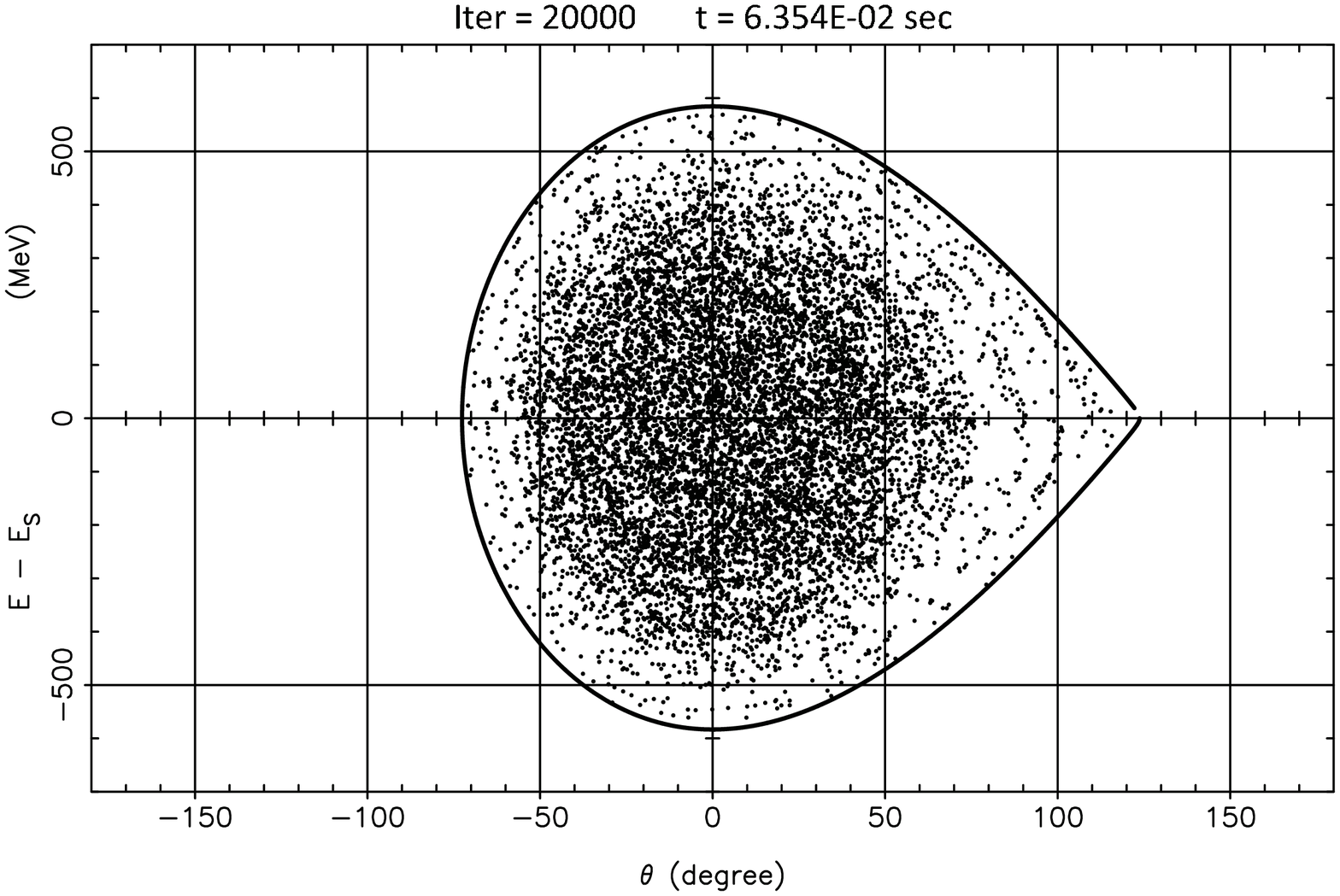}
\\ \centering \footnotesize (d)   \\
\includegraphics[width=4cm]{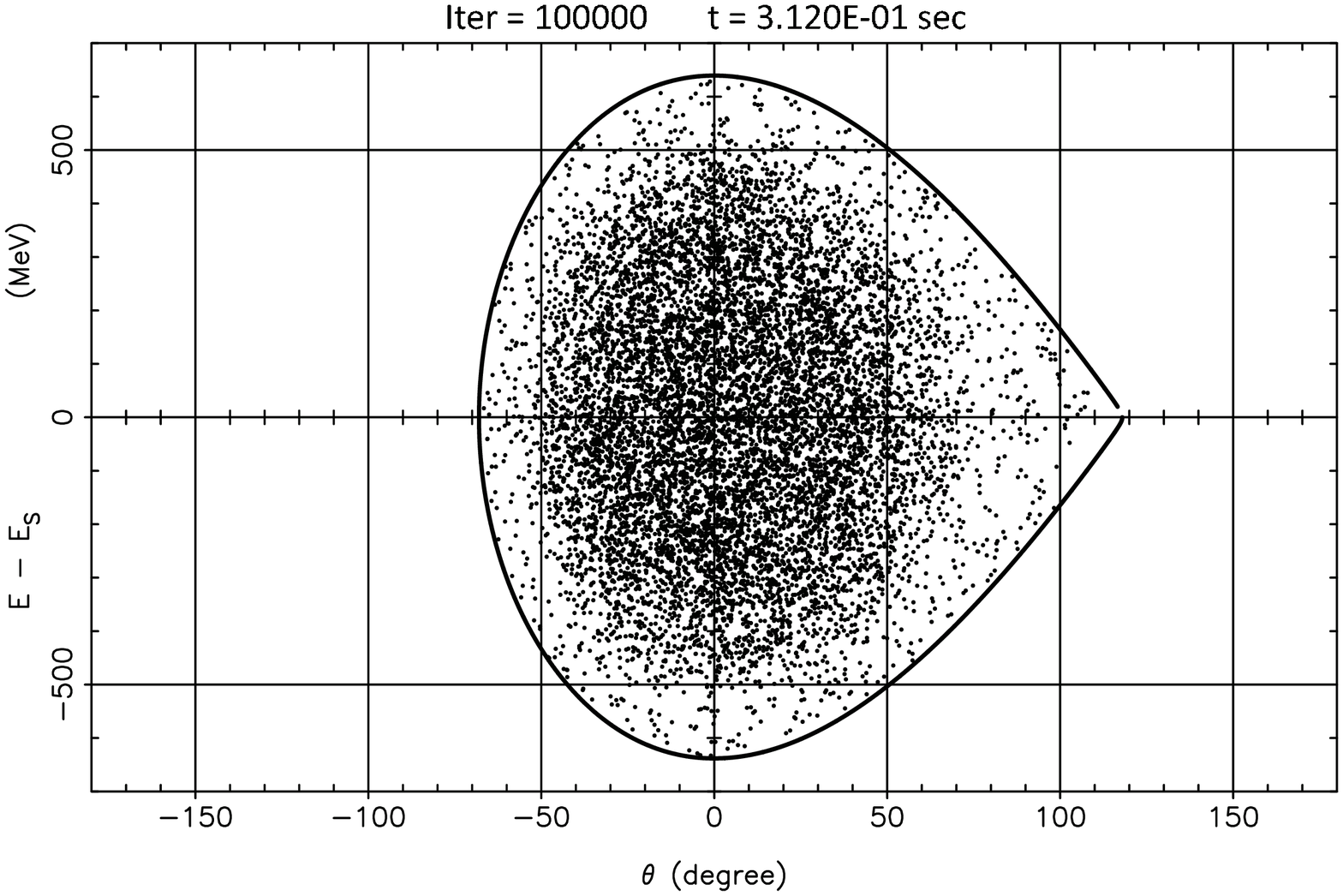}
\\ \centering \footnotesize (e)   \\
\includegraphics[width=4cm]{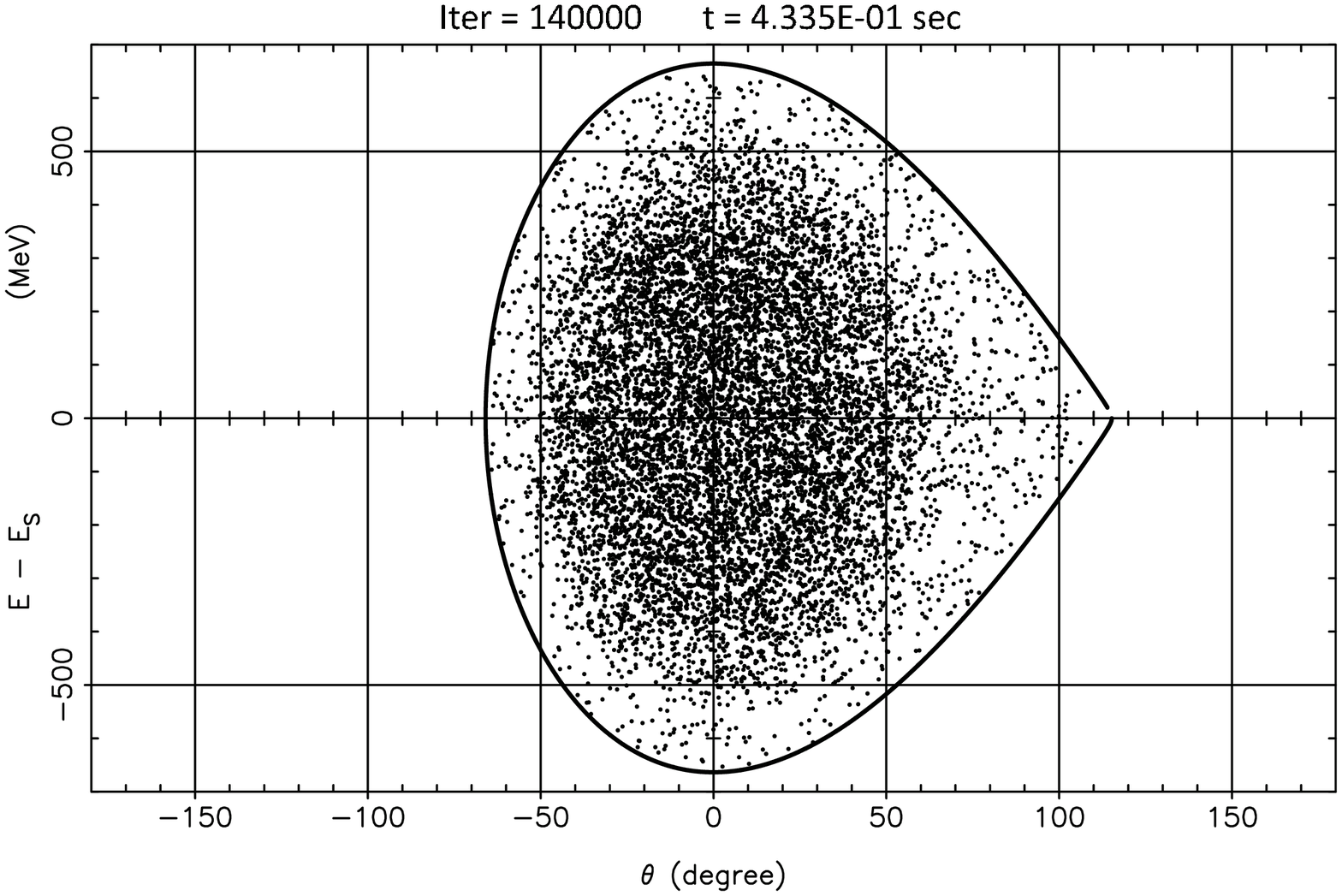}
\\ \centering \footnotesize (f)   \\

\end{multicols}
\figcaption{ Particle distribution of acceleration process using ESME program in 0th turn, 5000th turn, 10000th turn, 20000th turn, 100000th turn and 140000th turn.}
\label{acc_process}
\end{center}

At the end of acceleration process, we got a particle distribution as shown in Fig.~\ref{acc_result}.
\begin{center}
\centering
\includegraphics[width=7cm]{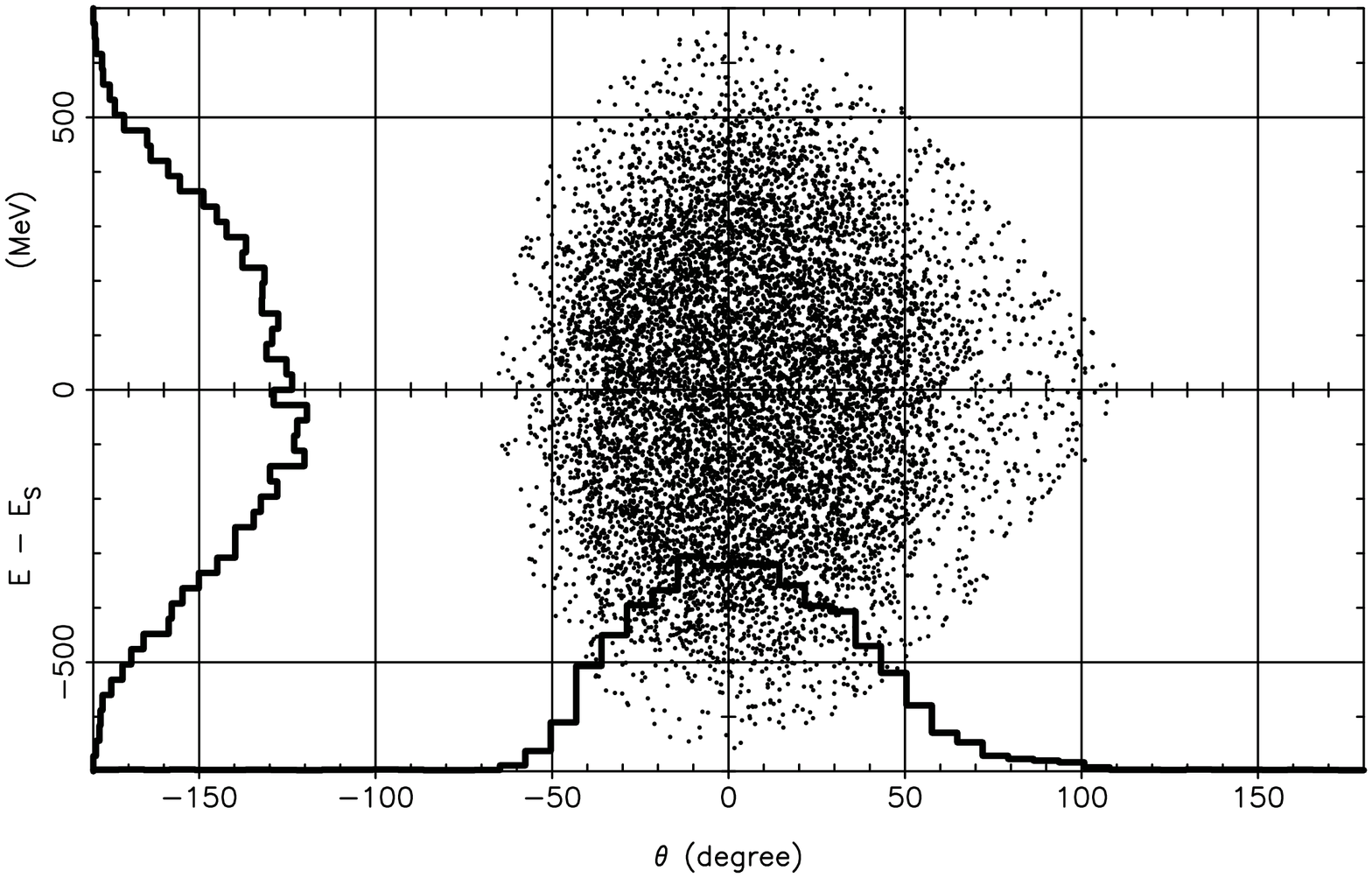}
\figcaption{Acceleration process result in the end.}
\label{acc_result}
\end{center}

\section{Conclusion}
We have used the code ESME to simulate the capture process and the acceleration process for $U^{34+}$ beam. The beam and machine parameters are expected for beam compression of heavy ion beams in CRing.

We have shown that capture efficiency can be 100\% theoretically with adiabatic capture program, and a 99.3\% acceleration efficiency can be achieved using the optimum programs we have mentioned.

Simultaneously, we have shown that the maximum RF voltage we used is 37.7 kV, and the peak voltage of the RF cavity in HIAF-CRing is designed to 40 kV, which can satisfy the RF requirement in both capture and acceleration processes.

Future work will concentrate on space charge effect to high current beams during capture and acceleration process. More precise results will be carried out considering the beam collective effects.

\end{multicols}

\vspace{-1mm}
\centerline{\rule{80mm}{0.1pt}}
\vspace{2mm}

\begin{multicols}{2}

\end{multicols}

\clearpage


\begin{thebibliography}{90}

\vspace{3mm}

\bibitem{lab1}YANG J C, XIA J W, XIAO G Q, XU H S, ZHAO H W, ZHOU X H, et al. Nuclear Instruments and Methods in Physics Research Section B: Beam Interactions with Materials and Atoms, 2013, {\bf 317}: 263---265

\bibitem{lab2} J. MacLachlan, J.-F. Ostiguy. User¡¯s Guide to ESME es2011\_4.5. 2011.

\bibitem{lab3} S.Y. LEE. Accelerator Physics. 2nd Ed.  Singapore, World Scientific, 1999. 242---243

\bibitem{lab4} Ng, K. Y. Adiabatic capture and debunching. Fermi National Accelerator Laboratory, Batavia, IL (United States), 2012.

\bibitem{lab5} Crescenti, M., P. Knaus, and S. Rossi. The RF Cycle of the PIMMS Medical Synchrotron. No. CERN-PS-2000-032-DR. 2000.

\bibitem{lab6} Ng K Y. PHYSICAL REVIEW SPECIAL TOPICS-ACCELERATORS AND BEAMS, 2002, {\bf 5}: 061002.

\bibitem{lab7} H¨¹lsmann P, Boine-Frankenheim O, Klingbeil H, Schreiber G: Considerations concerning the RF System of the accelerator chain SIS12/18-SIS100 for the FAIR-project at GSI. internal note 2004.

\bibitem{lab8} LIU W, XIA J W, ZHANG W ZH, LIU Y, YIN X J, WU J X, MAO L J, ZHOU X M, et al. High Power Laser and Particle Beams, 2005, {\bf 17}: 943---946
\bibitem{lab9} S.Y. LEE. Accelerator Physics. 2nd Ed.  Singapore, World Scientific, 1999. 247


\end{thebibliography}
\end{document}